\documentclass[aps,prl,twocolumn,superscriptaddress,longbibliography]{revtex4-1}

\usepackage{siunitx}
\usepackage[utf8]{inputenc}
\usepackage{graphicx}
\usepackage{graphics}
\usepackage{dsfont}
\usepackage{bm}
\usepackage{amsmath}
\usepackage{amsfonts}
\usepackage{mathrsfs}
\usepackage{braket}
\usepackage[capitalize]{cleveref}
\usepackage{units}
\usepackage{color}

\begin{document}
\title{Optoelectronic Reciprocity in Hot Carrier Solar Cells with Ideal Energy Selective Contacts}
\author{Andreas Pusch}
\email{a.pusch@unsw.edu.au}
\affiliation{School of Photovoltaic \& Renewable Engineering, UNSW Sydney, Kensington 2052, Australia}
\author{Milos Dubajic}
\affiliation{School of Photovoltaic \& Renewable Engineering, UNSW Sydney, Kensington 2052, Australia}
\author{Michael P. Nielsen}
\affiliation{School of Photovoltaic \& Renewable Engineering, UNSW Sydney, Kensington 2052, Australia}
\author{Gavin J. Conibeer}
\affiliation{School of Photovoltaic \& Renewable Engineering, UNSW Sydney, Kensington 2052, Australia}
\author{Stephen P. Bremner}
\affiliation{School of Photovoltaic \& Renewable Engineering, UNSW Sydney, Kensington 2052, Australia}
\author{Nicholas J. Ekins-Daukes}
\affiliation{School of Photovoltaic \& Renewable Engineering, UNSW Sydney, Kensington 2052, Australia}

\begin{abstract}
    Hot carrier solar cells promise theoretical power conversion efficiencies far beyond the single junction limit.
    However, practical implementations of hot carrier solar cells have lagged far behind those theoretical predictions.
    Reciprocity relations for electro-luminescence from conventional single junction solar cells have been extremely successful in driving their efficiency ever closer to the theoretical limits.
    In this work, we discuss how the signatures of a functioning hot carrier device should manifest experimentally in  electro-luminescence and dark $I-V$ characteristics.
    Hot carrier properties lead to deviations from the Shockley diode equation that is typical for conventional single junction solar cells. These deviations are directly linked to an increase in temperature of the carriers and therefore the temperature measured from electro-luminescence spectra.
    We also elucidate how the behaviour of hot carrier solar cells in the dark depends on whether Auger processes play a significant role, revealing a stark contrast between the regime of negligible Auger recombination (carrier conservation model) and dominant Auger recombination (Impact Ionization model) for hot carrier solar cells.
\end{abstract}

\maketitle

\section{Introduction}

One of the major limitations for the efficiency with which photovoltaic cells convert sunlight into electricity is the thermalization of electronic carriers with the crystal lattice of the solar absorber material~\cite{Hirst2011,Green2016a}.
Photons with energy higher than the band gap of the absorbing semiconductor lose their excess energy to this thermalization.
It was recognised several decades ago~\cite{Ross1982} that a solar energy converter in which electrons do not thermalize with the lattice could act as a heat engine with an efficiency limited by the Carnot efficiency corresponding to the temperature differential between the carriers and the lattice.
The efficiency of such a device could potentially reach $\approx 85\%$ \cite{Wurfel1997} at full concentration of sunlight if contacts with vanishing energetic bandwidth, but nonetheless high conductivity, could be achieved.

Extensive experimental evidence dating back several decades exists that demonstrates a transient difference between the electronic carrier temperature and the lattice temperature~\cite{ShahBook} lasting a few ps.  Several of these early spectroscopic results have been repeated in the context of photovoltaic hot carrier devices using III-V semiconductors, \cite{Clady2012} and later in various semiconductor nanostructures \cite{Hirst2013,Esmaielpour2018}.
InN has unsusal vibronic properties that has resulted in transient temperature differences of a few hundred ps \cite{Chen2003}.  Graphene photodetectors have been demonstrated whose photoresponse is dominated by hot electron transport \cite{Gabor2011,Tielrooji2013}.
and the cooling dynamics of hot carriers in graphene can be tuned by doping and by applying gate voltages \cite{Song2015}.
Recently, many different perovskite materials have been investigated for hot carrier properties \cite{Hopper2018,Tamming2019}. Cooling of hot carriers, on the order of tens of picoseconds, was observed with pump-probe experiments and attributed to the formation of polarons \cite{Joshi2019}, which, while protecting carriers from fast relaxation, also inhibit the energy redistribution between carriers that is crucial for efficient hot carrier solar cells \cite{Joshi2019,White2012}.
Theoretical models have been proposed to describe experimental observations by simulating the coupled electron phonon system after photo-excitation stimuli \cite{Hathwar2019} and consequent transport of excited particles \cite{Aeberhard2019}. 
Although hot carrier extraction has been experimentally demonstrated in III-V devices \cite{Dimmock2014,Hirst2014,Nguyen2018} the obtained efficiencies are still far away from improving upon single junction solar cells.

The concept of optoelectronic reciprocity that arises as a consequence of microscopic reversibility and detailed balance, has been used in single junction solar cells to infer device properties from electro-luminescence measurements.
It is usually written as \cite{Rau2007}
\begin{equation}
    \label{eq:optoRec}
    \phi_{em}(E_{\gamma},V) = EQE(E_{\gamma}) \phi_{bb}(E_{\gamma}) \Big( e^{\frac{q V}{k T_{c}}}-1 \Big) \, ,
\end{equation}
with the photon flux $\phi_{em}$ dependent on the photon energy $E_{\gamma}$ and the applied voltage $V$, according to the external quantum efficiency $EQE$ of the device, the blackbody spectrum $\phi_{bb}$ and the temperature of the device $T_{c}$.
This formulation is valid for single junction solar cells unless either the absorptivity or the current collection efficiency that together determine the external quantum efficiency change with applied voltage \cite{Aeberhard2017}.

In a hot carrier solar cell, reciprocity should take a different form since the current can be driven by both, temperature and electrochemical potential, differentials, making a hot carrier solar cell a hybrid between a photovoltaic and thermoelectric device.
For thermoelectrics, reciprocity is encapsulated in the direct relation to the Peltier and Seeback coefficients, which describe the heat carried by an electrical current and the electric field generated by a temperature difference, respectively \cite{NolasBook}.
This reciprocity is a direct consequence of the Onsager reciprocal relations \cite{Callen1948}.

To elucidate how optoelectronic reciprocity might manifest in hot carrier solar cells, we examine their properties when electronically driven under dark conditions.
Dark $I-V$ and electro-luminescence are powerful tools for the analysis of any solar cell architecture, yet the properties of hot carrier solar cells in the dark have not been discussed.
Here, we build a conceptual model for the I-V characteristics of a hot carrier cell with ideal energy selective contacts.
Special regimes in the $I-V$ curve depend on the importance of Auger processes in the description of the hot carrier solar cell but are always associated with an increase in carrier temperature that can be detected as a change in spectral shape of the electro-luminescence.

There are two competing models for the hot carrier absorber that consider two extremes regarding the importance of Auger recombination events.
The first is the carrier conservation model, developed by Ross and Nozik, which excludes Auger recombination and impact ionization processes and allows for a quasi-Fermi level separation (QFLS) in the absorber \cite{Ross1982}, and second is the impact ionization model, developed by W{\"u}rfel, that assumes a thermal hot carrier distribution \cite{Wurfel1997}, i.e. no QFLS, corresponding to infinitely fast Auger processes.
Note that the carrier conservation model is a better model for hot carrier absorbers with large band gaps while the impact ionisation model is more suitable for low band gap absorbers \cite{Takeda2009b,Wurfel2005}.

\section{Ideal energy selective contacts}

The goal for hot carrier solar cell contacts is to convert a temperature differential $\Delta T$ between the carriers in the absorber and the carriers in the contacts to an electrochemical potential difference, or voltage $V$, between the carriers in the two contacts.
In general, a current $I$ can be driven by electrochemical potential gradients and/or by temperature gradients \cite{NolasBook}
\begin{equation}
    \label{eq:generalTransport}
    I = q \int_{0}^{\infty} g(E) v(E) \tau_{e} \frac{\partial f_{0}(E)}{\partial E} \big( \frac{\partial \mu}{\partial x} + \frac{E-\mu}{T} \frac{\partial T}{\partial x}\big) dE \, .
\end{equation}
Here, $g$ denotes the density of electronic states, $v$ is the group velocity of electrons, $\tau_{e}$ is their relaxation time and $f_{0}$ is the Fermi-Dirac distribution, which depends on the spatially varying electrochemical potential $\mu$ and temperature $T$.

\begin{figure}
    \centering
    \includegraphics{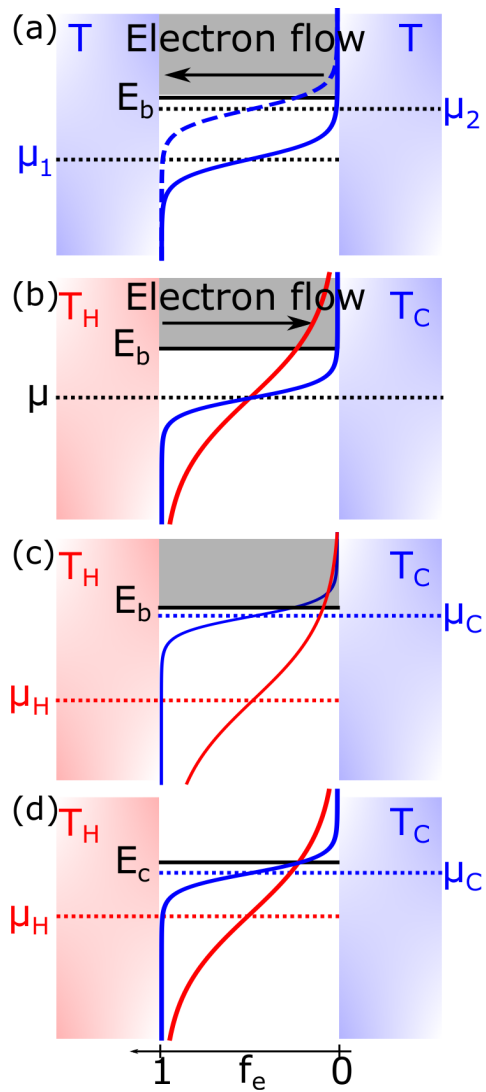}
    \caption{Electron transport from hot carrier absorber (right) to contact (left) across a thermionic barrier due to a difference in electrochemical potential (a) or a difference in carrier temperatures $T_{H}$ of the hot carriers on the left and $T_{C}$ of the cold carriers on the right (b) and the establishment of an equilibrium between electrochemical potential separation and temperature differential across the barrier (c). Carnot efficient conversion of a temperature differential into a electrochemical potential due to an infinitely narrow selective contact (d). The Fermi-Dirac functions $f_{e}$ show the distributions of electrons on the right (solid blue lines) and left (dashed blue/solid red lines) of the barrier/selective contact. Shaded areas indicate a non-vanishing density of states.}
    \label{fig:schottkyEnergySelective}
\end{figure}

To illustrate the transport equation we can first consider a contact, consisting of a thin semiconductor layer between two metal layers, that only allows carrier transport above a barrier energy, i.e. a Schottky or thermionic barrier.
Figure \ref{fig:schottkyEnergySelective}a depicts how a separation in electrochemical potentials generates a current across the barrier because of the presence of more carriers in the states above the barrier for the carrier population with a higher chemical potential.
A temperature differential also leads to an imbalance in carrier occupations above the barrier (see Figure~\ref{fig:schottkyEnergySelective}b) and a resulting flow of electrons.
This flow of electrons due to a temperature gradient can be compensated by an electrochemical potential gradient that acts in the opposite direction and an $I-V$ curve with positive electrical power production is traced out.
In open circuit the currents balance (see Figure~\ref{fig:schottkyEnergySelective}c).

Note that, under open circuit conditions, the net charge current will be zero across such a barrier, but a net energy flow persists as more low energy electrons flow from the higher electrochemical potential to the lower electrochemical potential while more high energy electrons flow from the high temperature region to the low temperature region.
This will lead to a reduction in efficiency of the electronic heat engine, which is well known in the context of thermoelectrics \cite{NolasBook}.
The dimensionless figure of merit $z T$ for a thermoelectric depends on the ratio of heat flows between short circuit and open circuit condition as $zT =\dot{Q}_{sc}/\dot{Q}_{oc}-1$~\cite{NolasBook}.

In a thermoelectric, electronic conduction competes with heat conduction through the lattice, so that there is inevitably a large contribution to the heat flow that is independent of the electro-chemical potential and dominates thermal conductivity at open circuit.
Therefore, one of the main design criteria for a thermoelectric is to increase the electronic conductivity compared to the thermal conductivity of the lattice.
In an ideal hot carrier solar cell, the heated carrier system in the absorber is insulated from the lattice of the absorber, so reducing the open circuit heat flow associated with electronic conduction while reducing the overall conductivity by limiting current to a small energy window can be a viable strategy to increase the efficiency of conversion of the heat gradient to electrical work.

Indeed, an optimal, isentropic \cite{Marti2013,Limpert2015b} conversion of a temperature differential into an electrochemical potential difference, i.e. a voltage, can be achieved if conduction is only possible in an infinitely narrow energy window, illustrated in Figure~\ref{fig:schottkyEnergySelective}d.

\begin{figure}
    \centering
    \includegraphics{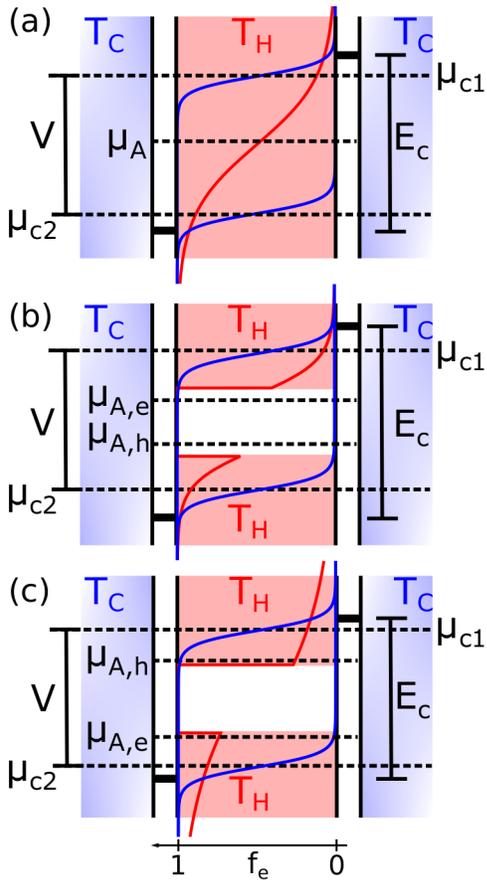}
    \caption{Schematic of the different hot carrier solar cell models with energy selective contacts and applied external voltage $V$. (a) The Impact Ionisation model for a hot carrier solar cell without band gap, (b) the carrier conservation model for a hot carrier solar cell with large band gap, with (b) positive and (c) negative QFLS in the absorber.
    Schematic Fermi-Dirac functions $f_{e}$ of the hot electron distribution in the absorber (red) and the cold electron distributions in the contacts (blue) illustrate that the occupation probabilities between contact and absorber are balanced at the energies of the respective ideal contacts, which leads to extraction at Carnot efficiency. Shaded areas indicate a non-vanishing density of states.}
    \label{fig:hotCarrierIllustration}
\end{figure}

The net current of the infinitely narrow contacts vanishes when
\begin{equation}
    \label{eq:carnotExtraction}
    \mu_{c1}-\mu_{A,e} \frac{T_{C}}{T_{H}} = E_{c1} (1-\frac{T_{C}}{T_{H}}) \, ,
\end{equation}
which can be rewritten in terms of the open circuit voltage $V_{oc}=\mu_{c1}-\mu_{c2}$ that balances carrier transport as \cite{Marti2013}
\begin{equation}
    \label{eq:carnotExtractionVoltage}
    V_{oc} = \Delta \mu_{H} \frac{T_{C}}{T_{H}} + E_{c} \big(1 - \frac{T_{C}}{T_{H}}\big) \, ,
\end{equation}
with the quasi-Fermi level separation (QFLS) in the absorber $\Delta \mu_{H}=\mu_{A,e}-\mu_{A,h}$ and $E_{c}= E_{c1}-E_{c2}$.
$V_{oc}$ represents the voltage for isentropic extraction \cite{Limpert2015b,Limpert2015} at the Carnot efficiency, visualised in Figure~\ref{fig:hotCarrierIllustration}.

This form for the open circuit voltage is only valid for ideal contacts.
Moving to a more realistic partially energy selective contact \cite{LeBris2010}, a thermionic barrier scheme \cite{Konovalov2019} or extracting from a side valley \cite{Ferry2019} introduces additional heat flows that complicate analysis.
Note that - contrary to claims made in \cite{Ferry2019} - equation (5) for the open circuit voltage in \cite{Ferry2019}, which is equivalent to equation \eqref{eq:carnotExtraction}, implicitly assumes an ideal energy selective contact to the L-valley and therefore Carnot efficient conversion of temperature differentials to voltage.
The actual open circuit voltage of the proposed device is thus overestimated. Neglecting the heat flow at open circuit also leads to a further overestimation of the absorber temperatures at open circuit.

If the contact conductivity $\sigma$ tends towards infinity, representing a delta-function contact~\cite{Mahan1996}, a finite current can be sustained at $V=V_{\text{balanced}}$.
Such an infinitely narrow contact with infinite conductivity is termed an ideal energy selective contact.
In the language of thermoelectrics, the figure of merit $z T$ of this ideal contact, combined with no heat exchange with the lattice, is infinite as there is no heat conduction at open circuit.
In a thermoelectric, the competition between heat conduction by the lattice with electronic conduction limits $z T$ to finite values even for such an ideal contact~\cite{Mahan1996,Humphrey2005}.

A real device will require a finite energetic window of conduction as an infinitely narrow window has zero conductivity.
The current for a window of finite energetic width $\Delta E$ can be written as
\begin{eqnarray}
    \label{eq:windowCurrent}
    I & = &  \int_{E_{c1}}^{E_{c1}+\Delta E}\sigma(E)\Big( \frac{1}{e^{\frac{E-\mu_{A,e}}{k T_{H}}}+1} - \frac{1}{e^{\frac{E-\mu_{c1}}{k T_{C}}}+1} \Big) dE \, \\ \nonumber
    & = &  \int_{E_{c2}}^{E_{c2}-\Delta E} \sigma(E) \Big( \frac{1}{e^{\frac{\mu_{A,h}-E}{k T_{H}}}+1} - \frac{1}{e^{\frac{\mu_{c2}-E}{k T_{C}}}+1} \Big) dE \, ,
\end{eqnarray}
with the energy dependent conductivity $\sigma(E)=v(E) g(E) \tau_{e}$, i.e. proportional to the number of conductive channels at a particular energy \cite{DattaBook}.
The $\mu_{ci}$ are the electrochemical potentials of electrons in contact $i$ and $\mu_{A,e/h}$ are the electrochemical potential of electrons and holes in the hot absorber with temperature $T_{H}$, while the $E_{ci}$ are the energies for the contact $i$ (illustrated in Figure \ref{fig:hotCarrierIllustration}).
Note that, for simplicity, we here assume that electron and hole density of states and therefore carrier temperatures are equal.

A finite energy width of the contact leads to an exchange of heat between absorber and contact also at open circuit as electrons with lower energies will flow predominantly from contact to absorber whereas electrons with higher energies flow predominantly from absorber to contact (compare Figure \ref{fig:schottkyEnergySelective}c).
Narrowing the bandwidth reduces the conductivity as fewer states are available for conduction, so that a compromise has to be found between conductivity and energy width \cite{Odwyer2008,LeBris2010,Limpert2015,Limpert2017}.
Practically, attempts at approaching such a contact have been made with resonant tunneling structures \cite{Dimmock2014,Dimmock2019}.

\subsection{Carrier cooling}

In order to consider carrier cooling we need to estimate the dependence of the heat transfer rate from carrier system to lattice on temperature and chemical potential of the hot carrier absorber.
This dependence is material dependent but we here build a model that fulfils sensible minimal conditions.
The heat transfer rate from the carrier population to the lattice $\dot{Q}$ should increase when the total number of carriers increases since the heat capacity increases also with the total number of carriers.
Therefore, it should scale with the intrinsic carrier concentration at the given temperature as well as with the QFLS in the absorber.
Assuming an undoped absorber, the QFLS is divided equally between electron and hole populations.
The heat transfer rate should also increase for higher temperatures in the absorber.
The simplest model that fulfills these criteria is given by
\begin{equation}
    \label{eq:heatTransferRateRN}
    \dot{Q} = \kappa e^{-E_{g}/(2 T_{H})} e^{\Delta \mu_{H}/(2 T_{H})} (T_{H}-T_{C}) \, ,
\end{equation}
with a proportionality constant $\kappa$.
This is a carrier-density dependent heat equation. 
$\kappa$ incorporates the density of states at the band gap but not the intrinsic carrier concentration due to temperature and band gap, which are captured separately in the two exponential factors in equation \eqref{eq:heatTransferRateRN}.
In the limit of vanishing band gap, we obtain $\dot{Q}(T_{H}) = \kappa (T_{H}-T_{c})$~\cite{Konovalov2019}, i.e. Fourier's heat conduction equation.

Note that, if heating is slowed because of hot phonon effects, the heat transfer rate will saturate for high carrier densities which is not included in our simplified cooling model.

\subsection{Impact ionization model}

In the Impact Ionization model of hot carrier solar cells, illustrated in Figure~\ref{fig:hotCarrierIllustration}(a), a total thermal equilibrium within the hot carrier population in the absorber is assumed.
This requires that energy redistribution by electron-hole scattering (i.e.~Auger recombination and impact ionization processes) occurs faster than energy is deposited in the system by photo-generation process of electrons and holes and extracted through the energetically narrow contact.

The main determinant of the timescale of Auger recombination and impact ionization compared to the timescale of electron-electron and hole-hole scattering is the size of the band gap compared to the temperature of carriers.
Auger recombination requires the presence of two electrons and one hole or two holes and one electron.
Thus the likelihood of the process depends exponentially on the ratio between band gap and temperature.
Impact ionization on the other hand requires a carrier with excess energy at least as large as the band gap, which again depends exponentially on the ratio between band gap and temperature.

Therefore, in the absence of a band gap, i.e. in metallic absorbers, no two distinct populations of carriers exist and the Impact Ionization model applies \cite{Wurfel1995}.
This model should also approximate a semiconductor where the band gap of the material is on the same order of magnitude or less as the thermal energy of the hot carriers \cite{Takeda2009b}.

In the absence of a quasi-Fermi level separation in the hot carrier system, equation \eqref{eq:carnotExtractionVoltage} fixes the temperature as a function of applied voltage at ideal energy selective contacts
\begin{equation}
    \label{eq:THfromV}
    T_{H}(V) = \frac{T_{c}}{1-\frac{V}{E_{c}}} \, .
\end{equation}
Here, $E_{c}=E_{c1}-E_{c2}$ is the energy difference between the ideal energy selective contacts.

Since the emission from the thermally equilibrated hot carrier system is determined by temperature and absorptivity of the hot carriers, this also means that electroluminescence properties as function of voltage are independent of carrier cooling and given by the Planck law \cite{Wurfel1982}
\begin{eqnarray}
    \label{eq:ELidealWuerfel}
    \phi_{em}(T_{H}) & = & \frac{2 \pi}{h^{3} c^{2}}\int_{0}^{\infty} \frac{\alpha(E) E^{2}}{e^{\frac{E}{T_{H}}}-1} dE \\ \nonumber
    & = & \frac{2 \pi}{h^{3} c^{2}}\int_{0}^{\infty} \frac{\alpha(E) E^{2}}{e^{\frac{E}{T_{c}}}e^{-\frac{E V}{T_{c} E_{c}}}-1} dE \, ,
\end{eqnarray}
with the energy dependent absorptivity $\alpha(E)$ and the carrier temperature $T_{H}$, determined from the voltage using equation \eqref{eq:THfromV}.
Note that the emission current is determined by the absorptivity, independently of whether the absorptivity is due to free carrier absorption or band to band transitions.
The impact ionization model describes a thermal emitter and only the energy balance, not the carrier balance, is important.

The temperature of the carriers is fixed by the applied voltage according to Eq.~\eqref{eq:THfromV}.
Thus carrier cooling does not reduce the temperature of the carriers in the presence of ideal energy selective contacts.
What then is its impact on the characteristics of the hot carrier solar cell in the dark?

\begin{figure}
    \centering
    \includegraphics{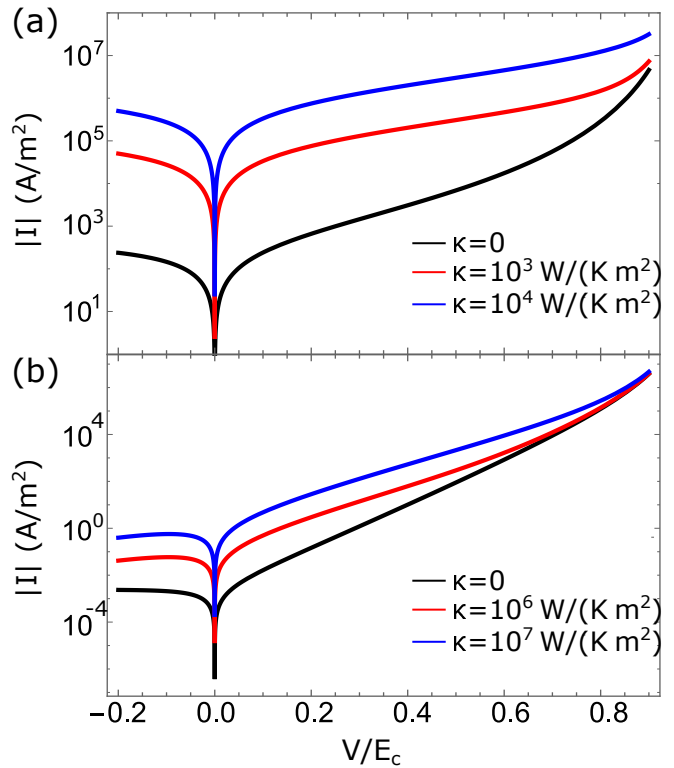}
    \caption{Dark $I-V$ characteristics (on logarithmic scale) of (a) a black body hot carrier solar cell calculated with the impact ionization model, assuming ideal energy selective contacts and with different cooling rates $\kappa$, corresponding to maximal, full concentration efficiencies of $70\%$ ($10^{3} \text{W}/(\text{K} \, \text{m}^{2})$) and $55\%$ ($10^{4} \text{W}/(\text{K} \, \text{m}^{2})$) \cite{Konovalov2019}, and (b) a hot carrier solar cell with band gap of $E_{g}=0.5$eV with different cooling rates. The voltage is normalised by the energy difference $E_{c}=1$eV between the two contacts.
    In the dark, the energy flow is always directed into the absorber for positive and negative bias.
    }
    \label{fig:IVblackbodyESC}
\end{figure}

Dark $I-V$ curves, assuming a perfect black body absorber, are shown in Figure \ref{fig:IVblackbodyESC} for different values of the cooling rate $\kappa$.

Since carrier cooling reduces the energy that is available to the hot carriers, the current that is required to maintain the temperature in the carrier system and the electro-luminescence dictated by the voltage applied at the contacts increases with increasing cooling rate.
This is analogous to the reciprocity relation in single junction solar cells where the voltage determines the electro-luminescence intensity but the current at a given voltage is also inversely proportional to the external luminescence extraction efficiency of the device.
A substantial, intermediate, exponential regime is found for large cooling rates but eventually the rise in current with voltage is super-exponential, corresponding to a rapid rise in temperature with the applied voltage.

The presence of a band gap reduces the overall emission rate and it also reduces the impact of thermal radiation from the environment on the $I-V$ curve, leading to an exponential regime for low voltages
(see Figure~\ref{fig:IVblackbodyESC}b) that also turns into a super-exponential regime for large voltages.
The impact of cooling on the $I-V$ curve is similar, yet the value of $\kappa$, at which an appreciable impact of cooling on the $I-V$ curve is observed, is much higher when a band gap is present.
This is due to the exponential dependence of the carrier density on the ratio between band gap and temperature.
At small voltages, i.e. small temperatures, carrier densities, and therefore heat transfer rates, are reduced by orders of magnitude compared to the case without band gap.

Note that in both, single junction and hot carrier cases (depicted in Figure~\ref{fig:IVblackbodyESC}), the interpretation of the $I-V$ curve is complicated by the presence of a finite series resistance $R_{s}$, which we have not included in our calculation since we assume an ideal contact.
In the presence of series resistance a larger current leads to a larger voltage drop across the device and therefore reduces the emission from the active region (in single junction solar cells) or the temperature of the hot carriers (in hot carrier solar cells).

Unlike the dark $I-V$ curve of a conventional solar cell, the $I-V$ curve of a hot carrier solar cell without a band gap (Figure~\ref{fig:IVblackbodyESC}) shows only a small exponential regime.

\subsection{Carrier number conservation model}

In the carrier conservation model for a hot carrier solar cell that was developed by Ross and Nozik \cite{Ross1982}, it was assumed that all absorbed photons would lead to exactly one conduction electron.
Conservation of the number of excited carriers in a semiconductor is an over-constraint and can be broken by, in principle avoidable, nonradiative recombination through defect states as well as unavoidable Auger recombination and impact ionisation.
Unlike defects, Auger recombination and impact ionisation are inevitable but their importance decreases exponentially with an increase in the ratio between the energy separation between quasi-Fermi levels and band edges and the temperature of the carriers.

To fulfill excited carrier number conservation and energy conservation simultaneously, a second degree of freedom in the Fermi-Dirac distribution of the carrier population is required.
Quasi-thermal equilibrium established within a population of electrons and a population of holes can be described by a quasi-Fermi level separation \cite{Wurfel1995}.
Thus it is a natural choice to allow for a quasi-Fermi level separation between electrons and holes in the absorber for hot carrier solar cells as well.
Because of the possibility of a cancellation between voltages associated with thermal gradients and electrochemical potential gradients, a QFLS in the absorber is even possible at short circuit, an effect that occurs in conventional single junction solar cells only to a small degree through finite mobility.

Allowing for a quasi-Fermi level separation $\Delta \mu_{H}$ complicates the relationship between applied voltage and temperature in the absorber as $\Delta \mu_{H}$ constitutes another independent variable \eqref{eq:carnotExtractionVoltage}.
It has been shown that this carrier number conservation model, in combination with ideal energy selective contacts, leads to unphysical predictions \cite{Wurfel2005}.
In fact, finite macroscopic differences in temperature and quasi-Fermi level separation result from infinitesimal changes in contact energy, when the contact energy crosses the mean energy of the absorbed photons.
This can be seen by inserting the equation for carrier number conservation
\begin{equation}
    \label{eq:carrierConservation}
    I = \dot{N}_{abs} - \dot{N}_{em} \, ,
\end{equation}
into the equation for energy conservation
\begin{eqnarray}
\label{eq:energyConservation}
I E_{c} = \langle E_{abs} \rangle \dot{N}_{abs} - \dot{N}_{em} \langle E_{em} \rangle \, ,
\end{eqnarray}
leading to
\begin{equation}
    \label{eq:Nratio}
    \frac{\dot{N}_{em}}{\dot{N}_{abs}} = \frac{\langle E_{abs} \rangle - E_{c}}{\langle E_{em}\rangle - E_{c}} \, .
\end{equation}
Equation \eqref{eq:Nratio} forces the temperature of the absorber to be higher than the mean temperature of the absorbed radiation when $E_{c}<\langle E_{abs} \rangle$ and smaller when $E_{c} > \langle E_{abs} \rangle$, creating a sharp cross-over point in absorber properties.
However, allowing for the inevitable presence of a small Auger interaction rate and a finite band width of the contact in real devices without compromising the essence of the model \cite{Takeda2009b} mitigates this conceptual problem.

It also implies that $\dot{N}_{em}$ can not be chosen to be arbitrarily small for any given combination of illumination conditions and contact energy.
As a consequence, the short circuit current obtained with this model may be substantially smaller than the absorbed photon flux.

The quasi-Fermi level separation $\Delta \mu_{H}$ in the absorber is a function of absorber temperature $T_{H}$, contact energy $E_{c}$, and applied voltage
\begin{equation}
    \label{eq:quasiFermiESCRossNozik}
    \Delta \mu_{H} = E_{c} - \frac{T_{H}}{T_{c}} (E_{c}-V) \, .
\end{equation}
The product $n p$ of the densities of electrons in the conduction band and holes in the valence band is proportional to $e^{\Delta \mu_{H}/T_{H}}$. 
An unusual, yet not unphysical prediction, of the carrrier number conservation model is the presence of temperatures in the absorber that are higher than the illumination temperature if the contact energy is smaller than the average energy of the absorbed photons.
This higher temperature is compensated for by a negative QFLS $\Delta \mu_{H}$ (see Figure \ref{fig:hotCarrierIllustration}c).
A negative QFLS leads to an exponential decrease in the $n p$ product and accordingly an exponential decrease in the emitted luminescence.
This phenomenon is called negative luminescence, i.e. less intense luminescence than expected for a thermal source with the same temperature and emissivity, and has been reported in various semiconducting materials under reverse bias \cite{Berdahl1989,Malyutenko2004}.
This way, the emitted photon flux will be lower than the incoming photon flux, despite the higher temperature in the absorber.
This net absorbed photon flux can be converted to electrical work if the bias at the contacts is positive.

The possibility of negative luminescence in hot carrier solar cells is related to the possibility of spontaneous negative luminescence and energy harvesting in reverse bias using a diode exposed to a cold environment, which also requires carrier number conservation.
This so-called thermoradiative diode~\cite{Strandberg2015b,Ono2019} is possible because it rejects proportionally more entropy than energy from the converter and the ultimate efficiency of the process is limited only by the Carnot efficiency~\cite{Pusch2019c}.
In contrast to the thermoradiative diode, however, the negative QFLS in the absorber region of a hot carrier solar cell with carrier number conservation is present despite a forward bias applied at the contacts.
The temperature differential between contacts and absorber allows for a turn-over of negative QFLS to positive bias.

For a hot carrier solar cell in the carrier conservation model, the emitted photon flux is given by
\begin{equation}
    \label{eq:ELidealRossNozik}
    \phi_{em}(V) = \frac{2 \pi}{h^{3} c^{2}}\int_{0}^{\infty} \frac{\alpha_{btb}(E) E^{2}}{e^{\frac{E-\Delta \mu_{H}}{k T_{H}}}-1} + \frac{\alpha_{fc}(E) E^{2}}{e^{\frac{E}{k T_{H}}}-1} dE \, .
\end{equation}
Here, we need to distinguish between the emission associated with the absorptivity $\alpha_{btb}$ related to band to band transitions and the emission associated with the absorptivity $\alpha_{fc}$ related to intraband free carrier absorption.
$\alpha_{fc}$ depends implicitly on the carrier density but the emission does not scale with the QFLS between electrons and holes as it occurs between carriers in the same band.
It contributes to the energy balance but does not contribute to the carrier balance, while $\alpha_{btb}$ also contributes to the carrier balance and formally replaces the EQE in the single junction reciprocity relation.

A frequency-dependent EQE, usually defined as the ratio of short circuit current to photon flux at a particular frequency, has no physical meaning in a hot carrier device due to its inherent nonlinearity.
The key assumption that validates equation \eqref{eq:ELidealRossNozik} is that the mobility in the absorber is infinite, so that the QFLS and temperature in the absorber are uniform.

\begin{figure}
    \centering
    \includegraphics{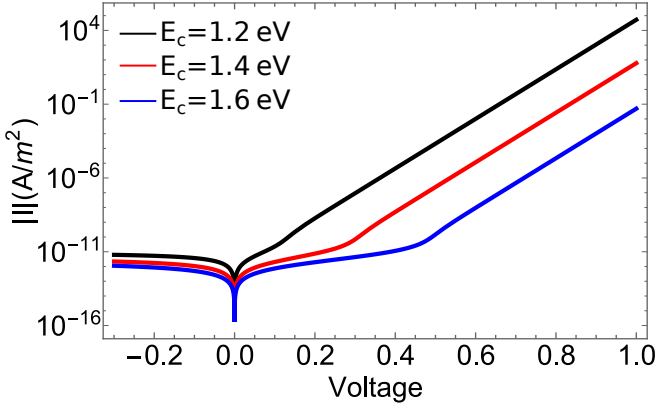}
    \caption{Dark $I-V$ characteristics (on logarithmic scale) for ideal hot carrier cells calculated with the carrier conservation model, assuming a band gap $E_{g}=1\,$eV and different contact energies $E_{c}$.}
    \label{fig:IVdarkRN}
\end{figure}

In the following, we neglect the free carrier contribution and assume that the absorptivity originates only from band to band transitions.
In this case, the photon flux can be integrated in Boltzmann approximation to
\begin{equation}
    \label{eq:ELidealBoltzmannPhoton}
    \phi_{em}(V) =  \frac{2 \pi}{h^{3} c^{2}} e^{\frac{\Delta \mu_{H}-E_{g}}{k T_{H}}} k T_{H} (E_{g}^{2}+2 k T_{H} E_{g} + 2 (k T_{H})^{2}) 
\end{equation}
with the mean energy per emitted photon given by
\begin{equation}
    \label{eq:ELmeanEnergyRN}
    \langle E_{em} \rangle = \frac{E_{g}^{3}+3 k T_{H} E_{g}^{2} + 6 (k T_{H})^{2} E_{g} + 6 (k T_{H})^{3}}{E_{g}^{2}+2 k T_{H} E_{g} + 2 (k T_{H})^{2}} \, .
\end{equation}
From the photon emission we can calculate the dark $I-V$ for hot carrier cells with carrier conservation
\begin{equation}
    \label{eq:darkIVRN}
    I(V) = \phi_{em}(V) - \phi_{abs} \, ,
\end{equation}
by considering only the absorption $\phi_{abs}$ of photons from the thermal photon bath of the environment.
For hot carrier cells with carrier conservation, shown in Figure \ref{fig:IVdarkRN} for the example of a band gap of $1\,$eV the ideal dark $I-V$ curves show an exponential increase over a wide range of biases, but diverge strongly from it at low bias.

\begin{figure}
    \centering
    \includegraphics{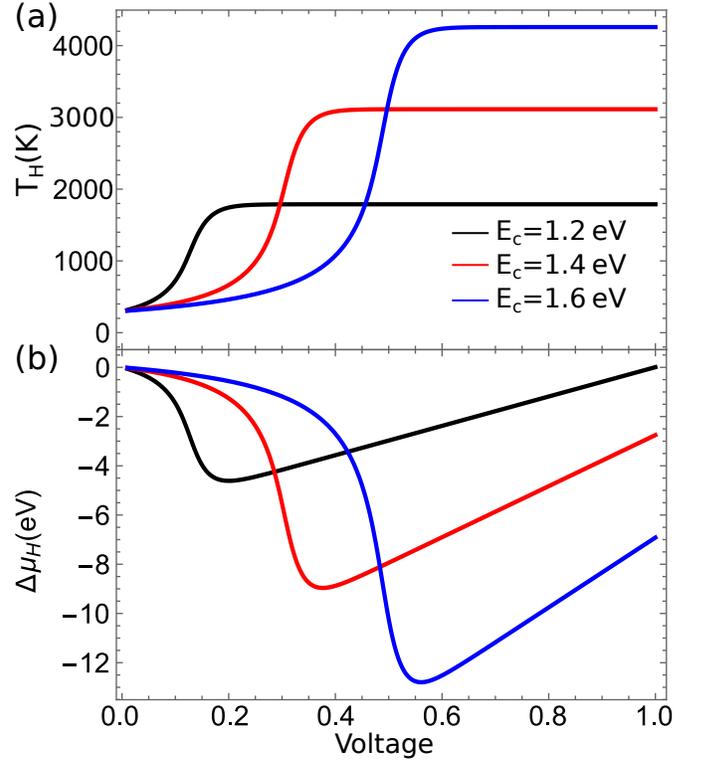}
    \caption{(a) Temperature and (b) QFLS in the absorber in the dark calculated with the carrier conservation model as a function of applied bias at the contacts for different contact energies and an absorber band gap of $1\,$eV.}
    \label{fig:tempMudarkRN}
\end{figure}

To understand this behaviour we need to look at the temperature and QFLS in the absorber as a function of bias, shown in Figure \ref{fig:tempMudarkRN}.
The carrier temperature of the absorber initially rises with bias and is then constant across a large bias regime in which absorption from the environment can be neglected compared to emission from the cell.
This is also the regime of the exponential behaviour in the $I-V$ curve.
In this regime the QFLS in the absorber changes at a rate proportional to the change in applied bias, i.e.
\begin{equation}
\label{eq:internalExternalBiasRN}
    \Delta \mu_{H}(V_{1})-\Delta \mu_{H}(V_{2}) = \frac{T_{H}}{T_{C}}(V_{1}-V_{2}) \, .
\end{equation}
Increasing contact energy increases the carrier temperature in the exponential regime, but leads to a delayed rise of carrier temperature with bias.

In the regime in which carrier temperature rises with bias, we observe a decrease in the QFLS between electrons and holes to negative values.
Thus, negative luminescence is predicted in this regime;
the temperature difference between absorber and contacts is compensated by a negative QFLS.
The onset of this regime and also the absolute value fo the QFLS reached depend on the contact energy.
The QFLS decreases more slowly from zero for higher contact energies, but decreases to a deeper minimum.

The Boltzmann approximation begins to break down at voltages near the contact energy $E_{c}$ and a break down of the Boltzmann approximation results in a deviation from the exponential behaviour.
Note, however, that the breakdown of the Boltzmann approximation is accompanied by substantial carrier populations in both the valence and conduction bands of the absorber.
The assumption of negligible Auger recombination is only valid for small carrier populations \cite{Takeda2009b}, so when the Boltzmann approximation breaks down the carrier conservation model also breaks down.

\begin{figure}
    \centering
    \includegraphics{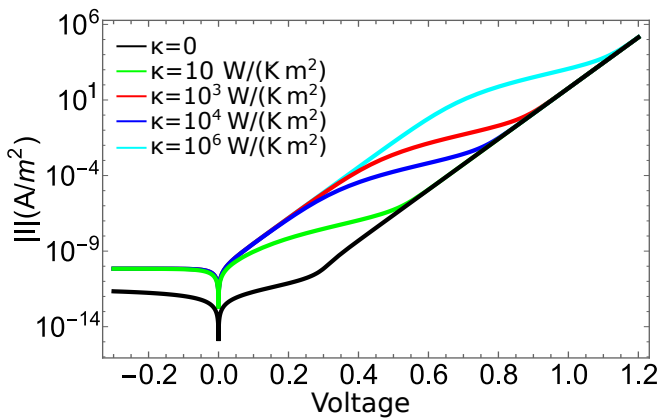}
    \caption{Dark $I-V$ characteristics (on logarithmic scale) for a hot carrier solar cell calculated with the carrier conservation model assuming a band gap of $1$eV and for different heat transfer rates.}
    \label{fig:heatTransferIVRN}
\end{figure}

Figure \ref{fig:heatTransferIVRN} shows $I-V$ characteristics for a hot carrier solar cell with a band gap of $E_{g}=1$eV, ideal selective contacts with $E_{c}=1.4$eV and several different cooling rates.
The presence of carrier cooling leads to an exponential regime for small biases, a cross-over regime where the dark $I-V$ deviates from exponential behaviour, and then again a region of exponential behaviour.
For larger cooling rates the cross-over regime is shifted towards ever larger bias, so that the dark $I-V$ approximates the behaviour seen from an ideal light emitting diode/solar cell over an ever increasing regime.
The hot carrier solar cell of the carrier conservation model with very large cooling rates is a conventional diode, described by Shockley's diode equation, with an unusual contact.

\begin{figure}
    \centering
    \includegraphics{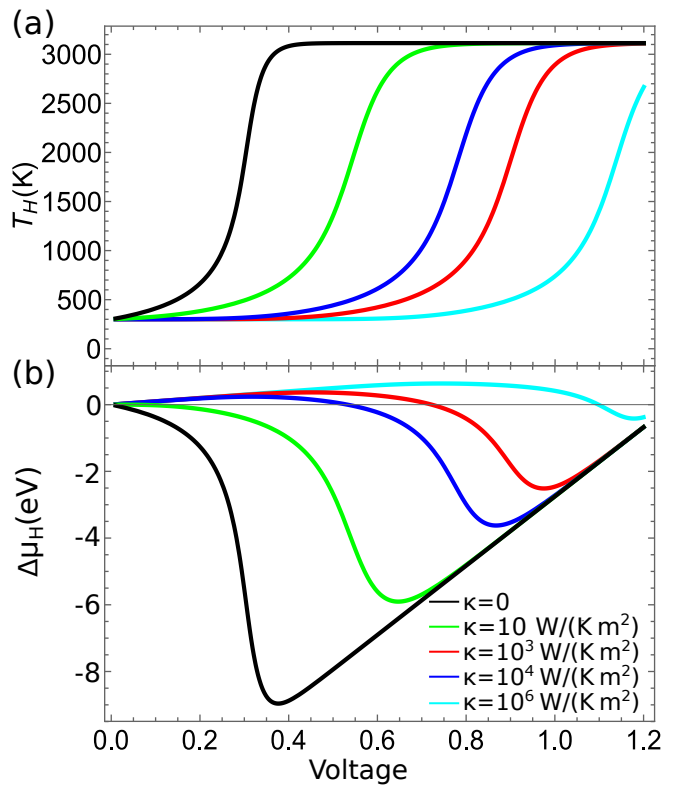}
    \caption{(a) Temperature and (b) QFLS in the absorber in the dark as a function of applied bias at the contacts, calculated with the carrier conservation model for different heat transfer rates $Q_{cc}$, an absorber band gap of $E_{g}=1$eV and a contact energy of $E_{c}=1.4$.}
    \label{fig:tempMudarkRNcooling}
\end{figure}

To understand the origin of the cross-over better we need to again examine the behaviour of temperature and QFLS in the absorber (see Figure \ref{fig:tempMudarkRNcooling}).
In the cross-over regime the temperature of the carriers in the absorber increases until it, surprisingly, reaches the same maximal value as in the absence of cooling.
This cross-over regime manifests in a change of spectral shape of the electroluminescence due to the increased carrier temperature.
The voltage at which this cross-over occurs is related to the cooling rate of the carriers.
The QFLS simultaneously decreases and, once the temperature saturates, it increases linearly with applied bias, leading to the second regime of exponential behaviour.

\begin{figure}
    \centering
    \includegraphics{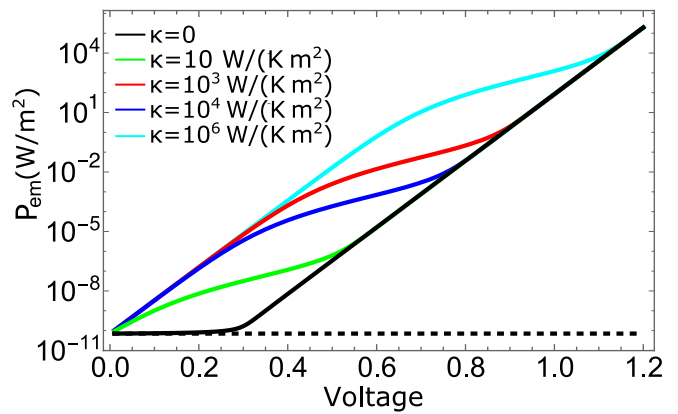}
    \caption{Total energy flux radiatively emitted from the hot carrier device (on logarithmic scale) as a function of applied bias in the dark for different cooling rates $Q_{cc}$. The dashed line indicates the energy absorbed from the environment.}
    \label{fig:emissionVsBias}
\end{figure}

The total energy flux radiatively emitted by the device can be calculated from the temperature and electrochemical potential distribution and is shown in Figure \ref{fig:emissionVsBias}.
Interestingly, the device without cooling shows the lowest radiative energy flux at low bias, which is due to the large negative bias in the absorber region.
The emission rate only starts to increase when the temperature has saturated.
This is because of the strict constraints in the carrier conservation model in which the total net emitted energy has to equal the net energy supplied by the current and the net photon flux also has to be the same as the current.
These strict requirements are lifted by the presence of cooling which provides another mechanism to balance energy and particle flows.
Consequently strong net emission is possible at any temperature, not only the temperature that equalises the mean emitted photon energy with the contact energy.

\section{Conclusions}

In this work, we presented the dark $I-V$ and electroluminescence of hot carrier solar cells with ideal energy selective contacts as tools for understanding and characterising hot carrier properties.

Low band gap absorbers with dominant Auger processes and ideally energy selective contacts show only a small exponential regime and the carrier temperature is determined by the ratio between applied voltage and contact energy.
Heat transfer to the lattice manifests as increase in current while the voltage dependence of the electro-luminescence properties is unaffected by the cooling rate.

Large band gap absorbers, for which Auger processes are slow, do show an exponential regime.
We underline that the prediction of a negative bias, that arises from the carrier conservation model with energy selective contacts, is not unphysical and results in negative luminescence.
In hot carrier absorbers with large band gap, the exact features of the deviations from the exponential regime depend on the heat transfer rate, allowing for the measurement of these rates from the dark $I-V$.
The carrier temperature is bias-, contact energy-, and heat transfer dependent but saturates at high bias and eventually becomes independent of heat transfer.

A finite Auger recombination and impact ionization rate, that interpolates between the two extreme models discussed in the paper, leads to a decrease in the absolute value of the QFLS compared to the prediction without Auger processes
This pushes the electronic system closer to an equilibrium between electrons and holes.
Auger processes also lead to a decrease in carrier temperature for the negative luminescence regime, because down-converting impact ionization dominates in that regime, whereas an increase in carrier temperature is expected in the positive QFLS regime, because up-converting Auger recombination dominates.
If realistic contacts with a finite width are considered, the results are expected to change qualititatively but the rise in carrier temperature, detectable as a change in spectral shape of the electroluminescence, is always associated with non-exponential (and non-ohmic) behaviour in the dark I-V.

\section{Acknowledgements}
This work was partially funded through the Australian Research Council Discovery Programme through project DP170102677.
M.P.N. thanks the UNSW Scientia Program for on-going support.

\bibliography{hotCarrierEL}

\end{document}